Review

# SANS evidence for the dispersion of nanoparticles in W–1Y$_2$O$_3$ and W–1La$_2$O$_3$ processed by hot isostatic pressing

A. Muñoz [a,*], J. Martínez [a], M.A. Monge [a], B. Savoini [a], R. Pareja [a], A. Radulescu [b]

[a] *Departamento de Física, Universidad Carlos III de Madrid, Avda. de la Universidad 30, 28911 Leganés, Spain*
[b] *Institut für Festkörperforschung, Neutron Science JCNS, Outstation at FRM II, 85747 Garching, Germany*

A B S T R A C T

The development of a dispersion of nanoparticles in the W–1%Y$_2$O$_3$ and W–1%La$_2$O$_3$ (wt%) alloys processed by hot isostatic pressing have been investigated using small-angle neutron scattering. The analyses of the scattering data using the Beaucage unified approach reveal the presence of a bi-modal distribution of spherical scattering centers with sizes of less of 180 nm in these alloys. The mode values of these centers are found at ~10 and 40 nm in W–1%Y$_2$O$_3$, and at ~15 and 80 nm in W–1%La$_2$O$_3$. The scanning electron microscopy analyses showed the presence of small second phase particles. The contribution of the pore space to the scattering curves has been analyzed using the results obtained for pure W processed following the same procedure used for the alloys, and the porosity measurements of the samples.

## Contents

## 1. Introduction

W and its alloys are very promising materials for making plasma facing components (PFCs) in the future fusion power reactors [1,2]. The properties that make W a suitable material for using in PFCs are its high melting point, good thermal conductivity, high thermal stress resistance, low tritium retention and high temperature strength. However, its high ductile–brittle transition temperature (DBTT) in the range 373–673 K, and recrystallization temperature (RCT) around 1500 K [3] limit the operating temperature range of those W components with structural functions. Nevertheless, the DBTT and RCT for W can be improved by addition of some dopants. Recently, W–1Y$_2$O$_3$ and W–1La$_2$O$_3$ alloys have been produced by mechanical alloying and subsequent consolidation by hot isostatic pressing [4,5]. The improvement of the high temperature mechanical properties of the oxide dispersion strengthened (ODS) W alloys would be associated with the fine dispersion of the oxide nano-particles, which can inhibit the dislocation motion and grain growth. A relative enhancement of the mechanical characteristics of the W–0.5%Y$_2$O$_3$ and W–(2–4)%Ti–0.5%Y$_2$O$_3$ alloys with respect to W and W–4%Ti have been reported [6,7]. This was attributed to the significant reduction of the oxidation rate observed in the alloys containing Y$_2$O$_3$ rather than to dispersion strengthening since the presence of a dispersion of nano-particles in those W alloys was not reported because their microstructural characteristics were not thoroughly investigated. Therefore, it would be very useful to verify the formation of such a dispersion of particles in oxide doped W and study their morphology and size distribution.

The small angle neutron scattering (SANS) technique, compared with others as transmission electron microscopy (TEM), is probably the most appropriate technique to analyze quantitatively the size distribution and morphology of the dispersoids in a matrix since it samples a significant volume of the material. In the present paper, the results of SANS measurements performed on the powder metallurgy W–1%Y$_2$O$_3$ and W–1%La$_2$O$_3$ (wt%) alloys are reported.

## 2. Experimental procedure

The W–1Y$_2$O$_3$ and W–1La$_2$O$_3$ alloys were produced by mechanical alloying and subsequent hot isostatic pressing for 2 h at 1573 K and 200 MPa following the procedure described elsewhere [4]. An ingot of non-ODS W was also consolidated under identical conditions to be used as reference material. The density of the materials was measured using a He ultrapycnometer. High resolution scanning electron microscopy (SEM) images were obtained in a MEB JEOL J8M6500 field emission scanning electron microscope equipped with an energy dispersive X-ray spectrometer (EDS).

The SANS experiments were carried out in the FRM II research reactor at Garching (Germany). The experimental data for the W–1Y$_2$O$_3$ and W–1La$_2$O$_3$ alloys were acquired using the KWS-1 diffractometer with a neutron wavelength of 7 Å and sample-detector distances of 2, 8 and 20 m. The SANS measurements for non-ODS W were made in the KWS-2 diffractometer using sample-detector distances of 2, 8 and 20 m and wavelengths of 3.5, 7.6 and 19.6 Å, respectively. These setups let a range of scattering vectors $Q$ between 0.002 and 0.14 Å$^{-1}$ be probed, which corresponds to scattering angles $\theta$ in the range 0.06–4.5° and scattering center sizes between 5 and 310 nm, approximately. The measured scattered intensities from the alloys were corrected by subtracting background counts and taking into account the detector efficiency. The resolution function of the instrument was taken into account in the fitting procedure of the experimental data. These intensities were calibrated in absolute units using a Plexiglas sample as secondary reference standard to obtain the scattering cross section value $\Delta\Sigma/\Delta\Omega$ from the scattering centers embedded in the matrix.

## 3. Results and discussion

The density values measured for the consolidated materials, along with the estimated porosity, are given in Table 1. The closed porosity, i.e. the volume fractions of closed pores, for the ODS W alloys and non-ODS W were respectively calculated by means of the following formulae:

$$P_{ODS} = 1 - \left(\frac{1-f_{ox}}{\rho_W} + \frac{f_{ox}}{\rho_{ox}}\right)\rho \quad (1)$$

$$P_W = 1 - \frac{\rho}{\rho_W} \quad (2)$$

where $f_{ox}$ is the mass fraction of oxide added to the ODS alloy, Y$_2$O$_3$ or La$_2$O$_3$, $\rho_{ox}$ the density of the oxide, $\rho_W$ the density of pure W and $\rho$ the density measured. For comparison, Table 1 also includes an estimation of the density for the full-dense alloys, $\rho_{cal}$, calculated applying the mixture rule to the nominal composition of the alloys. The density values of 19.25, 5.03 and 5.96 g cm$^{-3}$ were used for W, Y$_2$O$_3$ and La$_2$O$_3$, respectively.

The scanning electron microscopy analyses revealed that these W alloys consisted of an ultrafine structure of W grains with large oxide pools refilling interstices between prior W particles. The average grain sizes estimated from high resolution SEM images resulted in ~0.3 μm. Furthermore, these images reveal the presence of nanoparticles dispersed in the W matrix, which are either embedded in the W grains or at the grain boundaries, as shown in Fig. 1. The EDS analyses of the disperse second phase particles showed a high content of O and Y, or La, although their precise composition and crystalline structure have not been yet determined. It should be noticed that transmission electron microscopy (TEM) evidence for dispersion of spherical nanoparticles in W–4V–1La$_2$O$_3$ prepared following the same route has been reported elsewhere [5].

SANS data are usually presented as normalized curves in which the scattered intensity $I$ is plotted as a function of the scattering vector, $Q$. Since the present W alloys can be considered like dilute systems with particles not correlated to one another, the Beaucage unified approach for dilute systems was applied to analyze the scattering curves [8,9]. According to this approach, in the case of a matrix with particles that scatter independently, the scattering curves may be approximated by the summation of independent exponential-power functions, i.e. as

$$I(Q) = \sum_{i=1}^{n}\left(G_i \exp\left(-\frac{Q^2 R_{Gi}^2}{3}\right) + B_i\left\{\frac{[erf(QR_{Gi}/\sqrt{6})]^3}{Q}\right\}^{P_i}\right) \quad (3)$$

where the $i$ index refers to a particular region of the scattering curve where the data fit a single exponential-power function, or unified scattering function, $G_i$ is the Guinier prefactor, $R_{Gi}$ the gyration radius, $B_i$ the Porod prefactor and $P_i$ the Porod exponent for the corresponding function.

Fig. 2 shows the scattered intensity from the alloys in the as-HIPed conditions over the range of $Q$ values investigated. The adjustment of the scattering data to Eq. (3) was very well accomplished in terms of three functions for the W–1Y$_2$O$_3$ alloy and two functions for the W–1La$_2$O$_3$ alloy revealing that the scattering objects sizes exhibit a tri-modal and bimodal distribution, respectively. Each adjusted function corresponds to a specific range of $Q$ values marked in the scattering curves. The parameters of the unified scattering functions are listed in Table 2. These functions has a Porod exponent of $P=4$ except for the one indexed as I for W–1Y$_2$O$_3$ that has $P=4.9$. The $P$ value is related with the structure of the scattering objects. $P=4$ indicates scattering from objects that have a sharp and smooth matrix/particle interface, like polydisperse spheres, while $P>4$ corresponds to objects with diffuse matrix/particle interfaces where the scattering length density varies continuously [8].

**Table 1**
Theoretical and experimental density of the different consolidated alloys. The volume fractions of Y$_2$O$_3$, La$_2$O$_3$ and pores have also been determined.

|  | Calculated density $\rho_{cal}$ (g cm$^{-3}$) | Measured density $\rho$ (g cm$^{-3}$) | Oxide fraction $\Phi$ | Closed porosity $P$ |
|---|---|---|---|---|
| W | 19.25 | 18.79 ± 0.01 | – | 0.024 |
| W–1%Y$_2$O$_3$ | 18.95 | 17.082 ± 0.004 | 0.038 | 0.087 |
| W–1%La$_2$O$_3$ | 18.83 | 17.06 ± 0.03 | 0.032 | 0.094 |

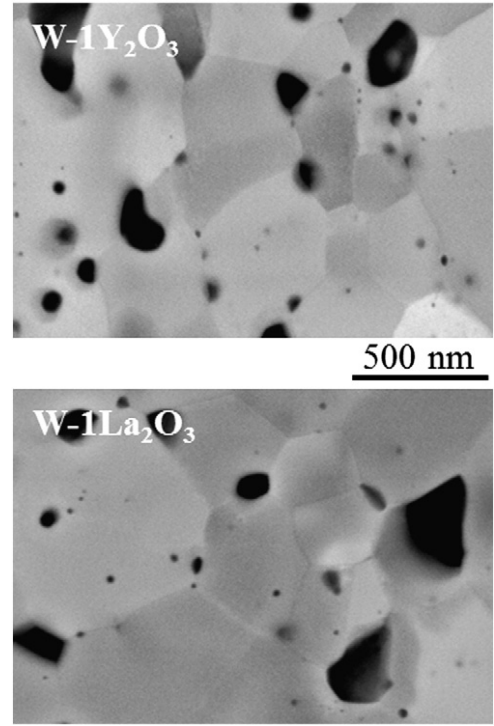

Fig. 1. High resolution SEM images showing Y-rich, or La-rich, particles in W–1Y$_2$O$_3$ and W–1La$_2$O$_3$. The presence of large particles at the interstices between grains is also evident.

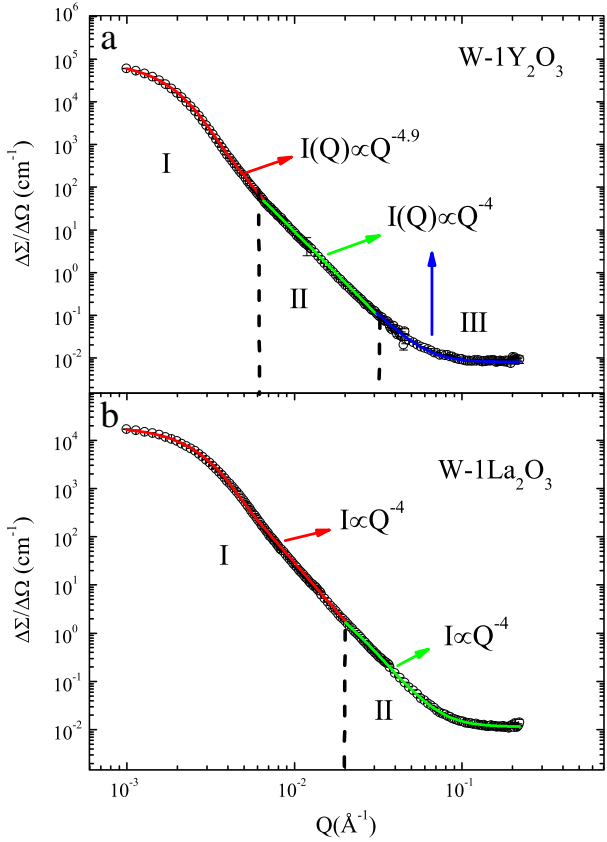

**Fig. 2.** Fits of the SANS data for W–1Y$_2$O$_3$ and W–1La$_2$O$_3$ to the empiric Beaucage unified model.

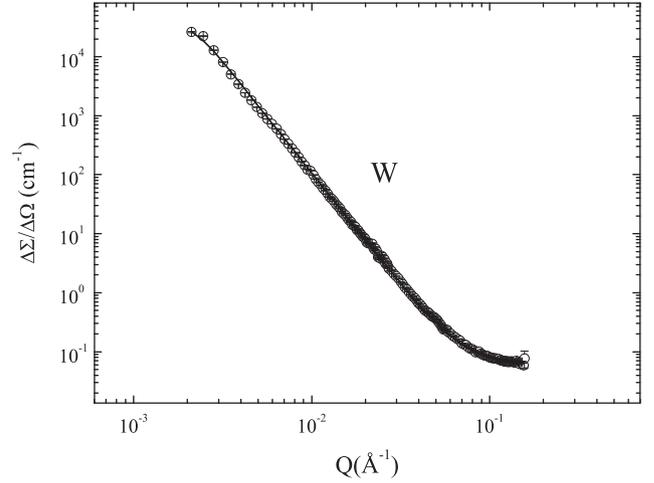

**Fig. 3.** Fits of the SANS data for pure W.

In contrast, SANS data for the non-ODS W material fit a single Beaucage function as Fig. 3 reveals. The corresponding adjustment parameters are also given in Table 2. The Porod exponent results in 3.8 indicating that these pores are spherical with a surface fractal structure of low roughness. In this case, scattering occurs exclusively at the interface between the W matrix and one type of scattering objects, which may only be pores since non-ODS W is chemically homogeneous. This result indicates that the pore size distribution detected is monomodal over the explored $Q$ range 0.002–0.14 Å$^{-1}$, i.e. in the pore size range 5–310 nm.

**Table 2**
Fitting parameters obtained from the SANS data analyzed using the empirical Beaucage approach, and mean radius $\langle R_i \rangle$ of the scattering centers.

| Material | Population index | $G_i$ cm$^{-1}$ | $B_i$ cm$^{-1}$ Å$^{-Pi}$ | $R_{Gi}$ (nm) | $P_i$ | $\langle R_i \rangle$ (nm) |
|---|---|---|---|---|---|---|
| W–1%Y$_2$O$_3$ | I | 9.4×10$^4$ | 1.0×10$^{-9}$ | 116±6 | 4.9 | |
| | II | 514.1 | 9.5×10$^{-8}$ | 44±1 | 4 | 55 |
| | III | 3.016 | 1.0×10$^{-7}$ | 13±1 | 4 | 16 |
| W–1%La$_2$O$_3$ | I | 1.9×10$^4$ | 2.9×10$^{-7}$ | 75±2 | 4 | 95 |
| | II | 23.0 | 3.5×10$^{-7}$ | 17±1 | 4 | 21 |
| W | I | 7.1×10$^{-6}$ | 2.95×10$^{-6}$ | 177±2 | 3.8 | 200 |

To obtain information on the scattering objects the SANS data of non-ODS W have been analyzed assuming that these materials are incompressible systems constituted by two-phases: a W matrix and the scattering centers embedded in the matrix. For a scattering structure there exists an invariant, $Z$, related to its scattering power that can yield model-independent information on the scattering contrast and volume fraction of scattering objects. This invariant $Z$ for a two-phase system would be given by [10]

$$Z = \int_0^\infty Q^2 I(Q) dQ = 2\pi^2 \varphi_2 (1-\varphi_2)(\rho_1^* - \rho_2^*)^2 \quad (4)$$

where $\varphi_2$ is the volume fraction of the phase 2, and $\rho_i$ the coherent scattering length density for the $i$-phase. The $Z$ value for non-ODS W has been evaluated from the SANS data shown in Fig. 3 by integration in the $Q$ range explored. An overall porosity of 0.028 is obtained using the scattering length densities of 3.06×10$^{10}$ cm$^{-2}$ for W and 0 for the pore space. This value is in good agreement with the closed porosity value of 0.024 estimated from the density measurements. It should be noted that neutrons probe open and closed pores, and submicrometer porosity can comprise closed or open pores, as well as networks of interconnected pores with arbitrary shapes. In addition, the $I(Q)$ curve for non-ODS W does not show an apparent trend to Q-independent scattering for values smaller than the lower bound of the experimental Q values, like the flattening of the $I(Q)$ curves of the ODS W alloys in Fig. 2 shows for the low Q values. This indicates that non-ODS W may have a significant number density of pores with linear sizes >310 nm. In contrast, the ODS alloys seem to have, in principle, a negligible number density of scattering objects with linear sizes >310 nm. It should also be noticed that the mean grain size of the ODS alloys is ~0.3 μm, and of the order of several microns for non-ODS W.

The mean particle radius $\langle R \rangle$ corresponding to the populations of scattering object with $P=4$, or 3.8, has been calculated assuming that each population can be represented by polydisperse spheres with a log-normal distribution of radii represented by a normalized number density function $f(R)$ given as [11]

$$f(R) = \frac{1}{\sqrt{2\pi}\sigma R} \exp\left\{-\frac{[\ln(R/m)]^2}{2\sigma^2}\right\} \quad (5)$$

where $R$ is the particle radius, $m$ the median radius and $\sigma$ a parameter related to the mean and median values of the radius by $\langle R \rangle = m \exp(\sigma^2/2)$. The $\sigma$ and $m$ values of $f(R)$ can be evaluated from the

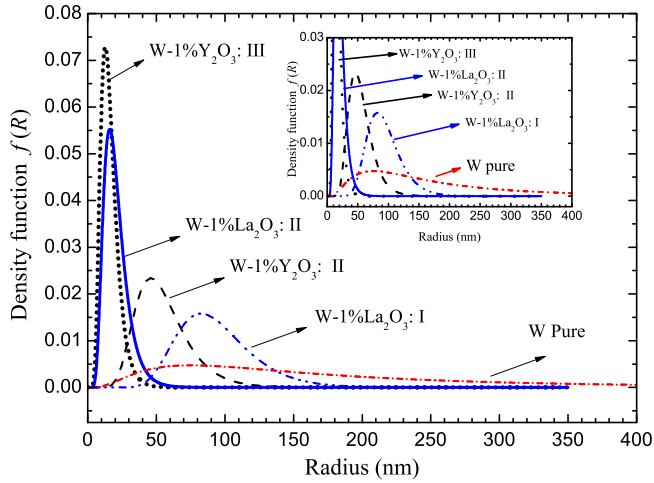

**Fig. 4.** Distributions of scattering center radii in W, W–1Y$_2$O$_3$ and W–1La$_2$O$_3$ obtained from the SANS data assuming polydisperse spherical centers.

fitting parameters obtained from the corresponding scattering data using the equations [11]:

$$\sigma = \left\{ -\frac{\ln\left[BR_G^4/1.62G\right]}{2\sigma^2} \right\}^{1/2} \quad (6)$$

and

$$m = \left\{ \frac{5R_G^2}{3\exp(1.4\sigma^2)} \right\}^{1/2} \quad (7)$$

The $\langle R \rangle$ values are given in Table 2, and the distribution functions $f(R)$ represented in Fig. 4.

The distribution function for population I with $P = 4.9$, detected in W–1Y$_2$O$_3$, was not attempted to obtain since the above approach is not applied for scattering objects yielding $P > 4$, i.e. for objects with diffuse matrix/particle interfaces [8]. Pores cannot be responsible for a scattering intensity with a Porod exponent of 4.9 because a matrix/pore interface with a scattering length density continuously changing is unfeasible. Neither are second phase particles embedded in the W matrix having interfaces with a steep change in the scattering length density. These scattering objects have to enclose an interface reaction zone in order to yield a continuous variation of the scattering length across the interface. This could be, tentatively, interpreted as evidence of Y$_2$O$_3$ decomposition or formation of complex oxides in pore space of larger dimensions.

Now, the nature of the objects giving scattered intensities described by a Porod exponent $P = 4$ in the ODS alloys has to be elucidated. At a first approximation, it can be assumed that the pore distribution in the ODS-W alloys is similar to that of the pure W, as they have been produced by following the same powder metallurgy route. According to the radius distributions represented in Fig. 4, the overlapping between the density function $f(R)$ for pores in non-ODS and the one for population III in W–1Y$_2$O$_3$, and for population II in W–1La$_2$O$_3$, suggests that the pore contribution to the scattered intensity may be assumed negligible in the Q range corresponding to populations III in W–1Y$_2$O$_3$ and II in W–1La$_2$O$_3$. Then, the scattering objects contributing to the intensity in these Q ranges may tentatively be attributed to second-phase nano-particles. On the contrary, the overlapping between the density functions for pores and populations II in W–1Y$_2$O$_3$ and I in W–1La$_2$O$_3$ indicates a noticeably contribution of the pores to the scattering in corresponding Q range. The radius of the scattering centers of these populations is found between ~30 and 120 nm. The SEM images shown in Fig. 1 reveal the presence of a significant number density of second-phase particles with these sizes. The presence of small pores in the sample surfaces is not evident in the high resolution SEM images but this does not prove the absence of inner small pores. The density of these pores still being small can significantly contribute to the scattered intensity because of the large scattering contrast.

## 4. Conclusions

The SANS data for the pure W and the ODS W–1Y$_2$O$_3$, W–1La$_2$O$_3$ alloys fit the Beaucage unified approach very satisfactorily. High-resolution SEM images show evidence of disperse second-phase particles in the matrix of the alloys. Their scattered intensity curves show the presence of a three-modal distribution of polydisperse spherical scattering objects in the ODS W–1Y$_2$O$_3$ alloy and a bi-modal distribution in the ODS W–1La$_2$O$_3$. In contrast, the scattering curve for pure W reveals the presence of a mono-modal distribution of spherical pores with matrix/pore interfaces that have a surface fractal structure of low roughness. The results indicate that the intensity scattered by populations III in W–1Y$_2$O$_3$ and II in W–1La$_2$O$_3$ are mostly attributable to second-phase particles with radii of less than ~60 nm, but scattering from pores may significantly contribute to the intensity scattered by the centers with radii in the range 60–310 nm.

## Acknowledgements

The present work was supported by the Consejería de Educación de la Comunidad de Madrid, through the program ESTRUMAT-CM S2009MAT-1585 and by the Ministerio de Innovación y Ciencia (project ENE2008-06403-C06-04). The additional financial support from EURATOM/CIEMAT association through contract EFDA WP11-MAT-WWALLOY is also acknowledged. The authors thank the Microscopy Laboratory of CENIM-CSIC for the high resolution SEM measurements.